\documentclass[
 reprint,
nofootinbib,
 amsmath,amssymb,
 aps,
pre
]{revtex4-2}

\usepackage{graphicx}
\usepackage{dcolumn}
\usepackage{bm}
\usepackage{xcolor}
\usepackage[normalem]{ulem}
\usepackage{mathtools}
\DeclarePairedDelimiter{\abs}{\lvert}{\rvert}       
\newcommand*\dif{\mathop{}\!\mathrm{d}}             

\usepackage[colorlinks]{hyperref}
\hypersetup{breaklinks=true}

\begin{document}

\preprint{...}

\title{Thermodynamic constraints and exact scaling exponents of flocking matter}

\author{Andrea Amoretti}%
\email{andrea.amoretti@ge.infn.it}
\author{Daniel K. Brattan}
\email{danny.brattan@gmail.com}%
\author{Luca Martinoia}%
\email{luca.martinoia@ge.infn.it}
\affiliation{%
Dipartimento di Fisica, Università di Genova, via Dodecaneso 33, I-16146, Genova, Italy \& \\
I.N.F.N. - Sezione di Genova, via Dodecaneso 33, I-16146, Genova, Italy.
}%

\date{\today}
\begin{abstract}
{\noindent We use advances in the formalism of boost agnostic passive fluids to constrain transport in polar active fluids, which are subsequently described by the Toner-Tu equations. Acknowledging that the system fundamentally breaks boost symmetry we compel what were previously entirely phenomenological parameters in the Toner-Tu model to satisfy precise relationships among themselves. Consequently, we propose a thermodynamic argument to determine the exact scalings of the transport coefficients under dynamical renormalisation group flow. These scalings perfectly agree with the results of recent state-of-the-art numerical simulation and experiments}.
\end{abstract}


\maketitle


\section{Introduction}
{\noindent Classical thermodynamics is the only physical theory of universal content that within the framework of applicability of its basic concepts will never be overthrown. Every challenger to the supremacy of this framework has fallen before it; whether they be black holes or, as we argue using advances in the theory of boost agnostic fluids, polar active systems.}

Active systems are often touted as out-of-equilibrium physics, where conventional equilibrium thermodynamic principles are expected to fall short \cite{Marchetti:2013}. These systems consist of self-propelled units, or active particles, that convert stored or ambient energy into directional movement. The interactions between active particles and their environment lead to highly coordinated collective motion and mechanical stresses. The continuous energy exchange with the surroundings and the intrinsic activity of the fluid's components drive these systems out of equilibrium, giving rise to unique behaviors such as pattern formation, non-equilibrium phase transitions and novel mechanical responses. These phenomena are prevalent in a wide array of systems ranging from bacteria to liquid crystals.

More specifically, a prominent role in this field is played by polar active systems, which are composed of self-propelled particles that align their movement with their closest neighbours. These systems demonstrate collective behaviors such as flocking and dynamic pattern formation, providing valuable insights into biological phenomena and serving as a framework for understanding complex systems.

Among the various theoretical approaches aimed at describing these systems, special attention is given to the model first proposed in \cite{Toner:1995,Toner:1998,Toner:2005,Toner:2012}. There, Toner and Tu put forward a phenomenological description of flocking behavior, drawing inspiration from the equations governing liquid crystals. They formulate a set of equations for a dynamic velocity field $\vec{v}$ and a conserved number density $n$, resembling the Navier-Stokes equation for passive fluids. The system's preferred background velocity is attained by coupling these equations to an external potential $U$ for the velocity field, akin to the well-known Mexican hat potential, which causes the system to undergo a spontaneous breaking of the rotational symmetry.

Unfortunately, despite the considerable success of the Toner-Tu model in elucidating the large-scale dynamics of polar active systems, the model itself lacks a clear microscopic derivation. This is a gap only partially addressed by kinetic theory derivations \cite{Bertin:2006,Bertin:2009,Ihle:2011,Beskaran:enhanced2008,Patelli:landau2021}, which however always rely on Galilean invariance in the form of Ward identities, as we shall discuss.

Returning to the Toner-Tu model itself, we remind the reader that the essence of any effective description is contained in identifying the relevant effective variables ($n$ and $\vec{v}$) and symmetries. In particular, we assume the system has translation invariance, which is only broken by the presence of the external potential $U$, but lacks a boost symmetry. Then, each velocity must be treated as a distinct configuration of the system, as opposed to Galilean or Lorentzian models where boosts relate systems at different velocities.

The formalism for describing fluids without boost invariance is “boost agnostic hydrodynamics” \cite{deBoer:2017ing,deBoer:2020xlc}. For such fluids, the velocity appears as a variable in the global thermodynamic description of the system – just like the temperature and chemical potential. The boost agnostic formalism, as per the name, does not discern between the mechanisms that lead to the velocity being necessary to specify equilibrium. Subsequently, we can add a constraint force – represented by $U$ - directly to our equations of motion to pick among the space of distinct velocities. As is typical in mechanics, such a constraint force has no consequences for the mechanical behaviour of the system, which in our case is the thermodynamic dependence of the transport coefficients. More importantly, adding such a force to the effective equations breaks translation invariance in a precise manner as one might expect from a theory with nearest neighbour interactions.

In this paper we present two novel results. First, we demonstrate that polar active fluids are akin to passive fluids lacking boost symmetry. This is opposed to some lore in the literature where the active fluid is completely divorced from the passive analogue. For hydrodynamics to offer a sensible description \cite{forster1975hydrodynamic}, we must always assume that there is a scale separation within the system, ensuring rapid local equilibration compared to long-scale collective dynamics.  The necessity of a local Gibbs-like distribution heavily constrains the constitutive relations of the ideal-fluid model. Notably, this implies that the quantity akin to pressure in the Toner-Tu equations can be treated as a thermodynamic pressure in many respects. Our argument centers on the requirement that an active fluid maintains consistency with basic thermodynamics, particularly in its disordered phase \cite{Agranov:2024}. While tuning the symmetry-breaking potential restricts the space of steady-state configurations and consequently affects collective dynamics and response to perturbations \cite{Toner:1995,Toner:1998,Toner:2005}, it does not alter the necessity for constitutive relations to adhere to thermodynamic principles. This argument is strongly supported by recent works, suggesting that standard hydrodynamics is a good description of systems that are in non-equilibrium steady states \cite{Amoretti:2022ovc,Brattan:2024}.

Secondly, these constraints enable us to derive analytical expressions for the critical exponents of the system in arbitrary spatial dimension $d$, aligning precisely with numerical computations in the Vicsek model \cite{Vicsek:1995} and differing from those derived by \cite{Toner:1998}. Our argument is grounded in the thermodynamic consistency of the ideal fluid constitutive relations, specifically the statement that at lowest order in derivatives the fluid should not produce entropy, and in principle can be applied to many different systems beyond the case of compressible polar fluids analyzed in this work.

\section{Thermodynamically consistent polar active fluids}
\noindent In this and the next section we will show that it is possible to recover all the observable features of the phenomenological Toner and Tu model from the hydrodynamics of a simple passive fluid.

As our starting point, consider a simple fluid without boost symmetry. The relevant hydrodynamic variables are the chemical potential $\mu$, associated with a conserved charge $n$, and the fluid velocity $\vec{v}$, which itself acts as a chemical potential for the fluid momentum $\vec{g}$. Following \cite{Armas:2020mpr,deBoer:2020xlc} we can define a generating functional for such fluid that produces the constitutive relations for the conserved currents in terms of a gradient expansion of the macroscopic variables. The leading term in a small derivative expansion for the effective action is then just the integral of a scalar, $P(\mu,\vec{v}^2)$, where $P$ is the usual thermodynamic pressure. Variations of the background on which the fluid sits produce the desired constitutive relations.  In particular, the ideal-fluid constitutive relations take the form \cite{deBoer:2017ing,deBoer:2020xlc,Armas:2020mpr,Amoretti:2022ovc}
	\begin{subequations}
	\label{eqn:idealTT}
	\begin{align}
		\label{eqn:ideal-fluid-constitutive-relations}
		g^i =\rho v^i \; , \; \; \tau^{ij}=\rho v^iv^j+P\delta^{ij} \; , \; \; j^i =n v^i \; , \; \;
	\end{align}
where $\vec{g} = (\partial P/\partial \vec{v})_\mu$ is the momentum conjugate to the velocity, $\rho = 2(\partial P/\partial \vec{v}^2)_\mu$ is the kinetic mass density and $n=(\partial P/\partial \mu)_{v^2}$. In the absence of external forces, they satisfy the following conservation equations
		\begin{align}
			\label{eqn:equations-of-motion}
			\partial_tg^i+\partial_j\tau^{ji}&=0 \; , \qquad \partial_tn+\partial_ij^i =0 \; .
		\end{align}
		\end{subequations}
The conservation laws of \eqref{eqn:equations-of-motion} are respectively associated with translation invariance and particle number. Notice in particular that\footnote{We set the particles' mass $m=1$ without loss of generality.} $\vec{j}\neq\vec{g}$, which is a consequence of the breaking of the Milne or Galilean Ward identity that holds only for systems with an exact boost symmetry \cite{Jensen:2014aia,Christensen:2013rfa}.

{\ The ideal constitutive relations \eqref{eqn:ideal-fluid-constitutive-relations} are the first terms in the derivative expansion corresponding to localising our macroscopic variables. The next terms in the derivative expansion, correcting $\tau^{ij}$ and $j^{i}$ in \eqref{eqn:idealTT}, lead to dissipative flows. This order-one hydrodynamics (for fluids without boost symmetry) has been studied in \cite{Novak:2019wqg,deBoer:2020xlc,Armas:2020mpr} and for ease of reference, we summarise the relevant expressions in the appendix \ref{appendix:A}. After obtaining the order one corrections, to complete the identification of our effective theory with the observed physics of polar active fluids, we turn on non-thermal noise and the external potential $U$ in the normative manner. In particular we add them to \eqref{eqn:equations-of-motion} by hand. In this regard we treat $U$ no differently to the noise $\vec{f}$ in assuming that the form of the constitutive relations are unchanged by these external forces. The result is that $\vec{g}$ is no longer a conserved quantity but it can nevertheless be a relevant hydrodynamic variable if the external potential, which breaks the symmetries, is weak enough \cite{Grozdanov:2018fic,Martinoia:2024cbw}. In this sense $U$ should be understood as imposing extra constraints on the system which reduce the allowed space of stationary states from those permitted to the passive fluid i.e. the stationary state velocity will be fixed to a particular value.}

{\ Following these considerations, we state the Toner-Tu equations:
\begin{subequations}\label{eqn:toner-tu-model}
	\begin{align}		&\partial_t\vec{v}+\lambda_1(\vec{v}\cdot\vec{\nabla})\vec{v}+\lambda_2(\vec{\nabla}\cdot\vec{v})\vec{v}+\lambda_3\vec{\nabla}\abs{\vec{v}}^2+\vec{\nabla}P_1\nonumber\\
		&\qquad\qquad+\vec{v}(\vec{v}\cdot\vec{\nabla}P_2)=U\vec{v}+D_B\vec{\nabla}(\vec{\nabla}\cdot\vec{v})\nonumber\\
		\label{eqn:toner-tu-model1}
		&\qquad\qquad\qquad+D_T\nabla^2\vec{v}+D_2(\vec{v}\cdot\vec{\nabla})^2\vec{v}+\vec{f} \; , \\
		&\partial_tn+\lambda_n\vec{\nabla}\cdot(n\vec{v})=0\label{eqn:toner-tu-model2} \; ,
	\end{align}
where all transport coefficients above are functions of $n$ and $\vec{v}^2$, $U$ is bounded below and has a zero at some $\vec{v}^2$ and we assume that $\vec{f}$ is a Gaussian white non-thermal noise
\begin{equation}\label{eqn:non-thermal-noise}
	\langle f_i(\vec{x},t)f_j(\vec{x}',t')\rangle=\Delta\delta_{ij}\delta^d(\vec{x}-\vec{x}')\delta(t-t')
\end{equation}
\end{subequations}
with $\Delta$ some constant. As was done in the original paper \cite{Toner:1998}, we have suppressed derivatives of the particle density in \eqref{eqn:toner-tu-model1} for legibility. These are easily restored and do not affect the physics that we will be interested in.}

{\ The Toner-Tu equations above \eqref{eqn:toner-tu-model} include $\lambda_{1}$, $\lambda_{2}$, $\lambda_{3}$ and $\lambda_{n}$ as phenomenological parameters. Our ideal charge conservation equation agrees with the Toner and Tu model \eqref{eqn:toner-tu-model} upon setting $\lambda_{n}=1$. At first glance, the same cannot be said about our ideal momentum equation, which from \eqref{eqn:ideal-fluid-constitutive-relations} we write as
\begin{equation}\label{eqnSM:momentum-conservation-ideal}
	\partial_t\vec{v}+\vec{v}(\vec{\nabla}\cdot\vec{v})+(\vec{v}\cdot\vec{\nabla})\vec{v}+\frac{\vec{\nabla}P}{\rho}+\frac{\vec{v}}{\rho}\partial_t\rho+\frac{\vec{v}}{\rho}\vec{v}\cdot\vec{\nabla}\rho=0 \ .
\end{equation}
Using the number conservation equation we can express the last two terms above as
\begin{align}
	  & \; \partial_t\rho+\vec{v}\cdot\vec\nabla\rho \nonumber\\
	&=-\left(\frac{\partial\rho}{\partial n}\right)_{v^2}n\vec\nabla\cdot\vec{v}+\left(\frac{\partial\rho}{\partial \vec{v}^2}\right)_n\left(\partial_t\vec{v}^2+\vec{v}\cdot\vec\nabla \vec{v}^2\right) \ .
\end{align}
Plugging this expression back into \eqref{eqnSM:momentum-conservation-ideal} and projecting the momentum equation along $\vec{v}$ we can solve for $\partial_t\vec{v}^2$. Finally, substituting the solution for $\partial_t\vec{v}^2$ again into the momentum conservation equation we arrive at the final result
\begin{align}\label{eqnSM:ideal-fluid-matching}
	& \partial_t\vec{v}+\lambda_1(\vec{v}\cdot\vec\nabla)\vec{v}+\lambda_2\vec{v}(\vec\nabla\cdot\vec{v})+\lambda_3\vec{\nabla}\abs{\vec{v}}^2+\frac{\vec\nabla P}{\rho} \nonumber \\
        & -k\frac{\vec{v}}{\rho^2}\bigl(\vec{v}\cdot\vec\nabla P\bigr)=0 \ ,
\end{align}
where the values of the parameters are
\begin{subequations}
        \label{Eq:ThermoTransport}
	\begin{align}
            \label{Eq:ThermoTransport1}
		\lambda_1&=1 \ ,	&	\lambda_2&=\frac{1-\frac{\partial\rho}{\partial n}\frac{n}{\rho}}{1+2\frac{\partial\rho}{\partial v^2}\frac{v^2}{\rho}}\ ,\\
            \label{Eq:ThermoTransport2}
		\lambda_3&=0\ ,	&		k&=2\frac{\partial\rho}{\partial v^2}\frac{1}{1+2\frac{\partial\rho}{\partial v^2}\frac{v^2}{\rho}}\ ,
	\end{align}
and the thermodynamic derivatives $\frac{\partial\rho}{\partial\vec{v}^2}$ and $\frac{\partial\rho}{\partial n}$ are to be understood at constant $n$ or $\vec{v}^2$ respectively. The last two terms in this equation \eqref{eqnSM:ideal-fluid-matching} can be identified as the transverse and longitudinal pressures $P_1$ and $P_2$:
	\begin{equation}
		\label{Eq:ThermoTransport3}
		\vec{\nabla} P_{1} = \frac{1}{\rho} \vec{\nabla} P \; , \; \;
		\vec{v} \cdot \vec{\nabla} P_{2} = \frac{\frac{2}{\rho}\frac{\partial\rho}{\partial v^2}}{1+2\frac{\partial\rho}{\partial v^2}\frac{v^2}{\rho}} \vec{v} \cdot  \vec{\nabla} P \; .
	\end{equation}
\end{subequations}
 we see that an anisotropic pressure term is a natural consequence of boost agnosticity even in equilibrium passive fluids.

Finally, we note that the coefficient $\lambda_{3}$ can be shifted to an arbitrary function. This happens because the pressure is a function of $\vec{v}^2$ and can be expanded as
\begin{equation}
	\frac{\vec\nabla P}{\rho}=\frac{1}{\rho}\left(\frac{\partial P}{\partial n}\right)_{v^2}\vec\nabla n+\frac{1}{\rho}\left(\frac{\partial P}{\partial\vec{v}^2}\right)_n\vec\nabla \abs{\vec{v}}^2 \ ,
\end{equation}
where the first term is interpreted as a new pressure gradient $\vec\nabla\tilde{P}$, while the second term defines $\lambda_3=\frac{1}{\rho}\left(\frac{\partial P}{\partial\vec{v}^2}\right)_n$. This was also the case in the original Toner-Tu model. We have made the logical choice for its value and assume that we are supplied with the form of the pressure $P$ to fix any ambiguity.}

{\ Restoring the boost symmetry amounts to setting $\vec{g} = n \vec{v}$ in \eqref{Eq:ThermoTransport} as is done in some works  \cite{Bertin:2006,Bertin:2009,Ihle:2011,Beskaran:enhanced2008,Patelli:landau2021}. This choice would force upon us the constraint $\lambda_{1}=1$ and $\lambda_{2}=\lambda_3=0$, as can be seen by direct substitution into \eqref{Eq:ThermoTransport1} and \eqref{Eq:ThermoTransport2}. Subsequently, one can redefine the chemical potential to remove the anisotropic pressure and general dependence of thermodynamics on $\vec{v}^2$. This conclusion is independent of the precise nature of the microscopic theory, following only from symmetries.}

{\ Interestingly, even without the imposition $\vec{g}= \vec{j}$, the requirement of having a local Gibbs-like distribution (i.e. a generating functional) has still heavily constrained the ideal fluid constitutive relations. Four a priori independent phenomenological parameters, $\lambda_{1}$, $\lambda_{2}$ and $\lambda_{3}$ and $\lambda_{n}$ from the original work \cite{Toner:1998} (see also \eqref{eqn:toner-tu-model}), are reduced to one ($P$) and its derivatives.}

\section{The ordered phase} 
\noindent Only for specific flows and choices of the potential $U$ in \eqref{eqn:toner-tu-model} would we expect the system of equations to arrive at a state with zero background velocity $\vec{v}=\vec{0}$ at late times. For a $\vec{v}^2$-dependent generic potential, bounded below by zero, the system will flow to a zero of $U$ and break rotations spontaneously. It is typical in the literature to take $U = \alpha - \beta \vec{v}^2$, so that $U \vec{v}$ behaves like a derivative of the Mexican hat potential. Around the new steady state we can then linearise in the velocity
\begin{equation}
	\vec{v}=(v_0+\delta v_\parallel)\hat{x}_\parallel+\vec{v}_\perp
\end{equation}
with $\delta v_\parallel$ and $\vec{v}_\perp$ small fluctuations which are respectively longitudinal or orthogonal to the background velocity  $v_0\hat x_\parallel$. $v_0$ is defined such that it obeys $U(n_0,v_0^2)=0$, with $n_0$ the equilibrium particle number density. 

{\ In the ordered phase the longitudinal mode associated with $\delta v_\parallel$ is gapped, therefore it can be integrated out to find a set of reduced equations for the Goldstone-like gapless mode $\vec{v}_\perp$ and $\delta n=n-n_0$. Our analysis is similar to that of \cite{Toner:2012}. In particular, we solve the equation for $\delta v_\parallel$ iteratively in terms of $\vec{v}_\perp$ and $\delta n$ treating derivatives and linearisation as small in the same parameter $\varepsilon$. On a practical level, this amounts to counting fluctuations and derivative on a similar footing $\mathcal{O}(\delta)\sim\mathcal{O}(\partial)\sim\mathcal{O}(\varepsilon)$, with $\varepsilon$ a counting parameter. The equation of motion for the heavy mode $\delta v_\parallel$, up to $\mathcal{O}(\varepsilon^2)$, reads
\begin{subequations}
\label{eqn:gapped-equation}
\begin{align}
        \label{Eq:HeavyMode}
	&\left(\rho_0+2v_0^2\rho_v\right)\partial_t\delta v_\parallel+\rho_0v_0\left(2\partial_\parallel\delta v_\parallel+\vec{\nabla}_\perp\cdot\vec{v}_\perp\right)+\nonumber\\
        &+\left(P_n+v_0^2\rho_n\right)\partial_\parallel\delta n+2v_0\partial_\parallel\delta v_\parallel\left(P_v+v_0^2\rho_v\right)+\nonumber\\
	&+v_0\rho_n\partial_t\delta n=\delta U(v_0+\delta v_\parallel)+\mathcal{O}(\varepsilon^3) \ ,
\end{align}
where $\rho_0=\rho(n_0,v_0^2)$ is the kinetic mass density computed on the background and we have introduced the notation
\begin{align}
	\left(\frac{\partial F}{\partial n}\right)_{v^2}&=F_n \ ,	&	\left(\frac{\partial F}{\partial\vec{v}^2}\right)_n&=F_v	\ ,&	\left(\frac{\partial^2 F}{\partial n^2}\right)_{v^2}&=F_{n^2} \ ,
\end{align}
with $F=\{U,P,\rho\}$. These thermodynamic derivatives are evaluated on the background at $n=n_0$ and $\vec{v}^2=v_0^2$ too. Non-linearities appear in \eqref{Eq:HeavyMode} from fluctuations of the external potential, which we expand as
\begin{align}
	\delta U&=U_n\delta n+U_v\left(2v_0\delta v_\parallel+(\delta v_\parallel)^2+\abs{\vec{v}}^2\right) \nonumber \\
                 & \; +\frac{1}{2}U_{n^2}(\delta n)^2+\mathcal{O}(\varepsilon^3) \ .
\end{align}
\end{subequations}

We can now solve equation \eqref{eqn:gapped-equation} iteratively for $\delta v_\parallel$ in terms of $\vec{v}_\perp$ and $\delta n$. At lowest order in fluctuations and derivatives the solution is simply given by
\begin{equation}\label{eqnSM:order-zero-solution}
	\delta v_\parallel\approx-\frac{U_n}{U_v}\frac{\delta n}{2v_0}+\mathcal{O}(\varepsilon^2) \ .
\end{equation}
We can subsequently plug this approximate solution back into the derivative and non-linear terms in \eqref{eqn:gapped-equation} and solve again for $\delta v_\parallel$. The solution takes the general form
\begin{align}\label{eqn:frozen_profile}
    \delta v_\parallel &=-\alpha_1\delta n-\alpha_2(\delta n)^2-\frac{\abs{\vec{v}_\perp}^2}{2v_0}-\alpha_3(\vec\nabla\cdot\vec{v}_\perp) \nonumber \\
    &\; -\alpha_4\partial_\parallel\delta n + \mathcal{O}(\varepsilon^3)
\end{align}
where the coefficients are 
\begin{subequations}
	\begin{align}
		\alpha_1&=\frac{1}{2v_0}\frac{U_n}{U_v} \ ,	\; \; \alpha_2=\frac{1}{8v_0^3}\frac{U_n^2}{U_v^2}+\frac{U_{n^2}}{4v_0}\ , \\
            \alpha_3&=\frac{\rho_n n_0}{2v_0U_v}-\frac{\rho_0}{2v_0U_v}-\frac{\rho_v n_0}{2v_0}\frac{U_n}{U_v^2}-\frac{n_0\rho_0}{4v_0^3}\frac{U_n}{U_v^2} \ ,\\
		\alpha_4&=\frac{\rho_0}{4v_0^2}\frac{U_n}{U_v^2}-\frac{U_n}{U_v^2}\frac{\rho_n n_0}{4v_0^2}-\frac{P_n}{2v_0^2U_{v^2}}+\frac{U_n}{U_v^2}\frac{P_v}{2v_0^2} \nonumber \\
                    &\; +\frac{U_n^2}{U_v^3}\frac{\rho_v n_0}{4v_0^2}+\frac{U_n^2}{U_v^3}\frac{n_0\rho_0}{8v_0^4}\ .
	\end{align}
\end{subequations}
Subsequently eliminating $\delta v_{\parallel}$ from the charge conservation equation we find, to $\mathcal{O}(\varepsilon^3)$,
\begin{subequations}
\label{eqn:linearisedTT}
\begin{multline}\label{eqn:charge-reduced-equation}
	\partial_t\delta n+v_2\partial_\parallel\delta n+n_0\vec{\nabla}_\perp\cdot\vec{v}_\perp+w_1\vec{\nabla}_\perp\left(\vec{v}_\perp\delta n\right)\\
	-w_2\partial_\parallel(\delta n)^2 -w_3\partial_\parallel\abs{\vec{v}_\perp}^2-\xi_n\partial_\parallel\left(\vec{\nabla}_\perp\cdot\vec{v}_\perp\right) \\ -D_\parallel\partial_\parallel^2\delta n-D_\perp\nabla^2_\perp\delta n=0
\end{multline}
where the precise expressions of the coefficients can be found in appendix \ref{appendix:B}. Similarly, the equation of motion for the transverse velocity fluctuations $\vec{v}_\perp$ becomes
\begin{align}\label{eqn:momentum-reduced-equation}
	&\partial_t\vec{v}_\perp+v_0\partial_\parallel\vec{v}_\perp+\kappa\vec{\nabla}_\perp\delta n+g_1\delta n\partial_\parallel\vec{v}_\perp+g_2\vec{v}_\perp\partial_\parallel\delta n\nonumber\\
	&+g_3\vec{\nabla}_\perp(\delta n)^2+g_4\vec{v}_\perp\bigl(\vec{\nabla}_\perp\cdot\vec{v}_\perp\bigr)+g_5\bigl(\vec{v}_\perp\cdot\vec{\nabla}_\perp\bigr)\vec{v}_\perp\nonumber\\
	&-D_v\vec{\nabla}_\perp\partial_\parallel\delta n-\xi_v\vec{\nabla}_\perp\bigl(\vec{\nabla}_\perp\cdot\vec{v}_\perp\bigr)\nonumber\\
	&-\xi_\perp\nabla^2_\perp\vec{v}_\perp-\xi_\parallel\partial_\parallel^2\vec{v}_\perp=\vec{f}_\perp \; . 
\end{align}
\end{subequations}
The $v_2$ and $\kappa$ terms are linear in fluctuations and derivatives, the $g_i$ and $w_i$ terms are the non-linear corrections, while the generalized diffusivities $D_i$ and viscosities $\xi_i$ are related to first-order transport coefficients, see appendix \ref{appendix:A}.

The form of these reduced equations, \eqref{eqn:charge-reduced-equation} and \eqref{eqn:momentum-reduced-equation}, matches exactly the expressions in \cite{Toner:2012}, with a few minor differences. First, we observe that we have a new non-linear term $g_4$, which is missing in \cite{Toner:2012}, secondly we find that the bare coefficient $D_\perp$ is non-zero, while it appears to be zero in \cite{Toner:2012}. Finally, $g_5=1$ and $v_0$ appears in front of $\partial_\parallel\vec{v}_\perp$, despite the lack of Galilean invariance. These modifications aside, we have fully reproduced the effective equations \eqref{eqn:linearisedTT} of Toner and Tu for the hydrodynamic modes starting from the hydrodynamics of a simple passive fluid. Thus the non-equilibrium features characterizing active matter are not relevant for the hydrodynamic description of transport, which we now confirm by examining the response of the system to perturbations. As the linearized equations are the same as in \cite{Toner:2012}, the modes and correlators take the same form, with different values of the coefficients.

In particular, one of the key signatures of the Toner and Tu model is the fact that there are two speeds of sound $v_{\pm}(\theta)$, where $\theta$ is the angle between the wave vector $\vec{q}$ and the background velocity. Specifically at $\theta=0$ one finds $v_+(\theta=0)=v_0$ and $v_-(\theta=0)=v_0\lambda_1$ \cite{Toner:1998}. This effect is observed in simulations \cite{Tu:1998} and experiments \cite{Geyer:2018}. Crucially, it was used to provide an estimate of $\lambda_1\approx0.75$, which was interpreted as a signature of the breaking of Galilean boost symmetry. As we have shown, even for boost-agnostic fluids we find $(g_5=)\lambda_1=1$. It would naively seem that our model fails to reproduce the observed phenomena - however this is not correct. Indeed, at $\theta=0$ we find that the sound speed once again has two values \cite{Toner:2012}
	\begin{align}
	\label{Eq:SpeedofSound}
		v_+(\theta=0)&=v_0	&	v_-(\theta=0)&=v_0\left(1-\frac{n_0U_n}{2v_0^2 U_v}\right)
	\end{align}
and the difference vanishes when $U_n=\left(\frac{\partial U}{\partial n}\right)_{v^2}=0$. The fact that $U_n\neq0$ is what allows us to have two different speeds of sound at $\theta=0$, despite having $\lambda_1=1$.

Interestingly, the parameter $U_n$ in \eqref{Eq:SpeedofSound} is key to the breaking of time-reversal invariance \cite{Amoretti:2023hpb,Amoretti:2023vhe}. Notice that the Toner and Tu equations \cite{Toner:1998} break Onsager relations \cite{Bowick:2022} not only in the coefficient $U_n$, but also through the $\lambda_i$ (see also \cite{Fodor:2016} for a discussion on time-reversal invariance in active-matter systems). This is very unusual from the point of view of hydrodynamics; usually in hydrodynamics the ideal fluid (which follows from local equilibrium) is always time-reversal invariant. Breaking of the Onsager relations only appears through viscous corrections or external forces. Contrary to this, the Toner and Tu model violates Onsager relations at lowest order in derivatives, therefore equilibrium (even in the disordered phase!) is not time-reversal invariant.

To emphasise the conclusions of this section: we are able to reproduce the phenomenology, for example the modes and response functions, of the Toner-Tu model from the hydrodynamics of passive fluids in the presence of an external potential while imposing all the thermodynamic relations one finds in the passive fluid. The active nature of the fluid does not need to modify the constitutive relations, in particular their thermodynamic character, to be compatible with observations. As we shall see in the next sections, this is further verified by considering the scaling behaviour of the transport terms under the dynamical renormalisation group.

\section{Dynamical Renormalization Group}
{\noindent Having derived the Toner-Tu equations from first principles we use the fact that several of our transport coefficients are pure functions of thermodynamic parameters, \eqref{Eq:ThermoTransport}, to our advantage; namely, we constrain the scaling behaviour of the transport coefficients. Using standard notations, we rescale fields and coordinates as
$(\vec{v}_\perp,\delta n,\vec{x}_\perp,x_\parallel,t)= (b^\chi\vec{v}'_\perp,b^\chi\delta n',b\vec{x}_\perp',b^\zeta x'_\parallel,b^zt')$. With this choice of scalings the equations of motion take the same form as before, but with rescaled transport coefficients given by:
\begin{subequations}\label{eqn:scaling-expressions}
	\begin{align}
		(w_1,g_{3,4,5})'&=b^{z-1+\chi}\left((w_1,g_{3,4,5})+\text{graphs}\right) \; , \\
		(w_{2,3},g_{1,2})'&=b^{z-\zeta+\chi}\left((w_{2,3},g_{1,2})+\text{graphs}\right)\; , \\
		(\xi_n,D_v)'&=b^{z-1-\zeta}\left((\xi_n,D_v)+\text{graphs}\right)\; , \\
		(\xi_\parallel,D_\parallel)'&=b^{z-2\zeta}\left((\xi_\parallel,D_\parallel)+\text{graphs}\right)\; , \\
		(\xi_{v,\perp},D_\perp)'&=b^{z-2}\left((\xi_{v,\perp},D_\perp)+\text{graphs}\right)\\
		\Delta'&=b^{z-\zeta+1-d-2\chi}\left(\Delta+\text{graphs}\right) \; ,
	\end{align}
\end{subequations}
where ``graphs'' represents perturbative corrections that are obtained from integrating out the fast modes with wave vector $b^{-1}\Lambda\leq \abs{\vec{q}_\perp}\leq\Lambda$ and $\Lambda$ is a UV cutoff \cite{Ma:1975,Forster:1977}. As was shown in \cite{Forster:1977,Toner:1998}, the asymptotic $\vec{q}\rightarrow0$ behaviour of the correlators is determined by the fixed point values of the scaling exponents $\zeta$, $z$ and $\chi$. Therefore, a lot of effort has been devoted to identifying the critical points of these $\beta$-functions using both clever symmetry arguments \cite{Toner:1998,Toner:2012birth,ikeda:2024scaling,ikeda:2024minimum} or through more direct computation \cite{Chen:2020,jentsch:2024new}.}

The simplest such fixed point example is the linear one, where non-linear terms are ignored. It is given by $(z,\zeta,\chi)=(2,1,\frac{2-d}{2})$. However, the linear fixed point is unstable for $d<4$, as non-linearities become relevant in the IR. Another example is found in \cite{Toner:1998}, where Toner and Tu obtained exact exponents in $d=2$ taking advantage of an emergent pseudo-Galilean symmetry. They obtain $(z,\zeta,\chi)=(6/5,3/5,-1/5)$ These results however are invalidated by the more accurate analysis of \cite{Toner:2012}, which agrees with our expressions \eqref{eqn:charge-reduced-equation} and \eqref{eqn:momentum-reduced-equation}, due to the presence of new non-linearities ignored in the original work.

\subsection{Entropy conservation and non-linear terms}
\noindent We now argue that we can fix the values of the critical exponents in \eqref{eqn:scaling-expressions} exactly}. Looking at the non-linear terms in our hydrodynamic expressions, the associated bare transport coefficients $g_i$ and $w_i$ all depend on global equilibrium thermodynamic properties (like $\rho_0$, $n_0$, $v_0$ and $P$), their thermodynamic derivatives (such as susceptibilities, compressibilities, \dots) and derivatives of the external potential $\left(\frac{\partial U}{\partial n}\right)_{v^2}$ and $\left(\frac{\partial U}{\partial \vec{v}^2}\right)_n$. As these quantities are thermodynamic in origin, they must not receive perturbative corrections as there are no hydrodynamic modes in global equilibrium, where these quantities are defined, that can renormalize these values.

More precisely, the hydrodynamic constitutive relations are not just the most general expressions compatible with symmetries; they are constrained by the second law of thermodynamics. For example, at first order in derivatives, the second law imposes well known constraints on the signs of the dissipative transport coefficients. Less appreciated is the fact that for ideal fluids the second law forces particular thermodynamic relations to be satisfied. To demonstrate this for the passive fluid, consider the entropy of the system, which is a function of the thermodynamic variables $s=s(n,\vec{v}^2)$, and its associated entropy flux $\vec{J}_s=\left(s-P/T_0\right)\vec{v}$ with $T_0$ some constant temperature. Then, using the thermodynamic relation $T_0\dif s=-\mu\dif n-\vec{v}\cdot\dif\vec{g}$, it can be shown that the the ideal fluid equations \eqref{eqn:idealTT} imply entropy is conserved $\partial_ts+\vec\nabla\cdot\vec{J}_s=0$. This result holds generically in hydrodynamics and tells us that the ideal fluid does not produce entropy as a consequence of local equilibrium. 

In the presence of an external potential $U$, the entropy of the ideal fluid satisfies a modified conservation law
\begin{equation}\label{eqn:entropy_conservation}
    T_0(\partial_ts+\vec\nabla\cdot\vec{J}_s)+U\vec{v}^2=0 \; . 
\end{equation}
We emphasise that this equation is a consequence of the ideal fluid equations of motion and the definition of the entropy $s$. That \eqref{eqn:entropy_conservation} holds identically for all solutions of the ideal fluid equations of motion is subsequently a consequence of thermodynamic relations between coefficients appearing in the other transport equations. As such, \eqref{eqn:entropy_conservation} holds for any solution of the fluid equations of motion including those in which the longitudinal fluctuations are integrated out using \eqref{eqn:frozen_profile}. This can be seen directly if we can expand \eqref{eqn:entropy_conservation} up to order two in fluctuations and eliminate $\delta v_\parallel$ using \eqref{eqn:frozen_profile} to obtain
\begin{align}\label{eqn:entropy_reduced}
    &T_0\delta_t\delta s+\varphi\vec\nabla\cdot\vec{v}_\perp+\pi\partial_\parallel\delta n+f_1\delta n(\vec\nabla\cdot\vec{v}_\perp)+\nonumber\\
    &+f_2(\vec{v}_\perp\cdot\vec\nabla)\delta n+f_3\partial_\parallel\abs{\vec{v}_\perp}^2  +f_4\partial_\parallel(\delta n)^2 +\mathcal{O}(\partial^2)=0 \; , 
\end{align}
where $\varphi$, $\pi$ and the $f_i$ are again functions that depend on the thermodynamics and the external potential, see the appendix \ref{appendix:B}. Applying the definition of $s$ in terms of the other charges, $T_0\partial_t\delta s=-\mu\partial_t\delta n-(v_0+\delta v_\parallel)\delta_tg_\parallel-\vec{v}_\perp\cdot\partial_t\vec{g}_\perp$, employing the reduced equations of motion \eqref{eqn:charge-reduced-equation} and \eqref{eqn:momentum-reduced-equation}, and also the reduced equation for $\partial_t g_\parallel$, we indeed find that \eqref{eqn:entropy_reduced} is identically satisfied\footnote{One needs to be careful as in \eqref{eqn:momentum-reduced-equation} the time derivative acts on velocity fluctuations, while to check \eqref{eqn:entropy_reduced} we need the time derivative to act on the momentum.}.

For the bare theory at the ideal level, satisfaction of \eqref{eqn:entropy_conservation} independently of the flow is just a consequence of thermodynamic relations between transport coefficients. Once we proceed with the first step of the dynamical renormalization group and integrate out the fast modes we will find that the equations \eqref{eqn:charge-reduced-equation}, \eqref{eqn:momentum-reduced-equation} and \eqref{eqn:entropy_reduced} have the same form, but for renormalized values of the transport coefficients. Our argument is that the renormalized dynamical equations \eqref{eqn:charge-reduced-equation} and \eqref{eqn:momentum-reduced-equation} should still identically solve \eqref{eqn:entropy_reduced} upon using thermodynamic relations.

To clarify this last step, we can look at a specific example. Inserting the equations of motion in \eqref{eqn:entropy_reduced} we find many terms, some of the simplest ones being
\begin{subequations}\label{eqn:constraint_example}
    \begin{align}
    &\left(f_3-\mu_0 w_3-\frac{1}{2}v_0\rho_0\right)\partial_\parallel\abs{\vec{v}_\perp}^2=0 \; , \\
    &\left(\frac{\partial\rho}{\partial n}v_0^2-2\frac{\partial\rho}{\partial v^2}v_0^3\alpha_1+\kappa-v_0\alpha_1\rho_0+\right.\nonumber\\
    &\qquad\qquad\qquad\qquad +f_2+\mu_0w_1\Bigr)\vec{v}_\perp\cdot\vec\nabla\delta n=0 \; , 
\end{align}
\end{subequations}
which, as expected, vanish for the bare values of the transport coefficients reported in appendix \ref{appendix:B}. Suppose that we perform the first step of the dynamical renormalization group on the equations \eqref{eqn:charge-reduced-equation}, \eqref{eqn:momentum-reduced-equation} and \eqref{eqn:entropy_reduced}. This in general will leave the form of the equations invariant, but will change the values of the transport coefficients due to graphical corrections e.g. $f_i\rightarrow f_i'=f_i+\delta f_i$. This process will not affect the coefficients in the longitudinal momentum equation, since the fluctuation $\delta v_\parallel$ is not dynamical. Subsequently, we find the same constraints as in \eqref{eqn:constraint_example}, but written in terms of the renormalized values $f_i'$ and $w_i'$. Finally, imposing that the the equation is still identically satisfied for the new values of the coefficients amounts to requiring that the perturbative corrections obey
\begin{align}\label{eqn:constraints}
    \delta f_2+\mu_0\delta w_1&=0    &   \delta f_3-\mu_0\delta w_3&=0 \; . 
\end{align}
Because the chemical potential $\mu_0$ is a non-universal function, and because it does not appear in our equations of motion, the only way to make the above expressions vanish is to have that $\delta f_2=\delta f_3=\delta w_1=\delta w_3=0$. At order two in fluctuations we find one last non-trivial relation $\delta f_4-\mu_0\delta w_2=0$, which implies that $w_2$ cannot receive perturbative corrections either. Therefore, the non-linear transport coefficients $w_1$, $w_2$ and $w_3$ in \eqref{eqn:charge-reduced-equation} cannot receive graphical corrections if the ideal fluid is not to produce entropy.

To find all the possible constraints one should expand the entropy equation \eqref{eqn:entropy_conservation} to order three in fluctuations, although the resulting expressions become rather long. Specifically, at order three in fluctuations we find the constraints
\begin{align}
	\delta g_i+\delta l_i&=0
\end{align}
where $l_i$ are coefficients that arise expanding the entropy equation \eqref{eqn:entropy_conservation} up to order three in fluctuations. These relations seem less restricting than those in \eqref{eqn:constraints}, however had we used $n$ and $\vec{g}$ as our variables of choice to write \eqref{eqn:momentum-reduced-equation} (which are the natural variables to study entropy conservation, instead of $n$ and $\vec{v}$), the above constraints would present thermodynamic derivatives $\partial\vec{v}/\partial\vec{g}$ in front of $\delta g_i$, suggesting again that the perturbative corrections of $g_i$ and $l_i$ must be independent, and thus they should all vanish.

The usual arguments used to forbid graphical renormalization are grounded in the continuous symmetries of the equations of motion (e.g. the Galilean symmetry in Navier-Stokes hydrodynamics \cite{Forster:1977}). Here we rely on another continuous symmetry, entropy (non-)conservation. This latter symmetry can be formalised in terms of KMS symmetries of the Schwinger-Keldysh path integral (see \cite{Armas:2024iuy} for a discussion). This argument, perhaps unknowingly, is what prompted Toner and Tu to assert that $\lambda_n$ in \eqref{eqn:toner-tu-model2} should be kept fixed at 1 \cite{Toner:1998}. Here, we propose that the same argument can be generalized to include all the ideal-fluid transport coefficients.

We remark that this conclusion is in agreement with all known exact results obtained in the past based on the pseudo-Galilean invariance. Specifically, results for Malthusian flocks in $d=2$ \cite{Toner:2012birth,Besse:2022} and $d=3$ \cite{Chen:2020moving,Chen:2020} and incompressible flocks in $d=3$ \cite{Chen:2018} all agree that non-linear terms should not receive perturbative corrections. The only exception are incompressible polar fluids in the disordered phase, for which perturbative corrections have been explicitly computed \cite{Chen:2015}. Nevertheless, even this last case can be understood from our argument. In the incompressible disordered phase, both chemical potentials $\mu_0$ and $v_0$ are zero, moreover $\delta v_\parallel$ cannot be integrated out. Hence the thermodynamics of the system simplifies significantly and we find $\delta l_i+\delta g_i=0$ without any thermodynamic quantity in front of $\delta g_i$, thus the only constraint is that perturbative corrections to the entropy equation transport coefficients must be equal to perturbative corrections to the momentum equation ones.

\subsection{Exact scaling exponents}
\noindent Having understood that the non-linear terms should not receive perturbative corrections, we can impose this on the scalings of the coefficients $g_i$ and $w_i$ in \eqref{eqn:charge-reduced-equation} and \eqref{eqn:momentum-reduced-equation}, finding two equations for the fixed point
\begin{align}
	z-1+\chi&=0	&	z-\zeta+\chi&=0 \; . 
\end{align}
To completely fix the exponents we need a third equation and as such we will employ the hyperscaling relation. This additional constraint follows from assuming that the scaling of $\Delta$ is exact. The reason for this is very practical, and comes from the fact that recent simulations suggest it is a universal law \cite{Mahault:2019}. Furthermore, a constant noise correlator $\Delta$ should in general receive contributions from at least two non-linearities, each carrying a power of $\vec{q}$. Corrections are then $\mathcal{O}(q^2)$ which are irrelevant compared to $\Delta$ in the hydrodynamic limit. Imposing this last condition
\begin{equation}
	z-\zeta+1-d-2\chi=0
\end{equation}
we find the exact scaling exponents
\begin{align}
        \label{Eq:Scalings}
	z&=\frac{2+d}{3}	&	\zeta&=1	&	\chi&=\frac{1-d}{3}
\end{align}
valid for all $2\leq d\leq4$. These expressions correctly interpolate to the mean field theory results which hold for $d\geq4$. They are in perfect agreement with recent state-of-the-art numerics on the Vicsek model \cite{Mahault:2019} which found discrepancies when comparing to the standard Toner and Tu expressions. These are outlined in Table \ref{table:exponents}. We also included in the table the large number fluctuation exponents $\langle\delta N\rangle^2\sim\langle N\rangle^\alpha$. These exponents should be 1 for thermal phases, while polar active matter hydrodynamics predicts $\alpha=1+(d+\zeta+2\chi-1)/d$ \cite{Ramaswamy:2003,Marchetti:2013,Mahault:2019}.  In the specific case of $d=2$ our argument also matches recent studies based on symmetry arguments \cite{chate2024}.

As expected, $\chi<0$ in $d=2$, implying that velocity fluctuations are weak and do not destroy the ordered phase. The Mermin-Wagner theorem would forbid continuous phase transitions below two dimensions, however active matter systems defy this expectation due to their ``non-equilibrium nature''. In the past this has been attributed to $\lambda_i$ terms in \eqref{eqn:toner-tu-model}. On the contrary, in our approach, the $\lambda_i$ are constrained by local equilibrium and do not meaningfully depend on any activity. Yet nevertheless, as we conclude this section, we find we have reproduced the necessary scaling properties of the transport coefficients. This adds to our evidence from previous sections where we recovered the transport pheonomenology of the active fluid.

\begin{table}
	\centering
	\begin{tabular}{c|ccc|ccc}
		&	\multicolumn{3}{c|}{$d=2$}	&	\multicolumn{3}{c}{$d=3$}\\
		&	TT98	& Vicsek	& this work	& TT98 & Vicsek  & this work\\
		\hline
		$\chi$		&		$-1/5$				&		$-0.31(2)$				&	$-1/3$ 		&	$-3/5$	&	$-0.62$	&	$-2/3$\\
		$\zeta$		&		$3/5$ 				&		$0.95(2)$				&	$1$ 		& 	$4/5$	&	$1$		&	$1$\\
		$z$			&		$6/5$ 				&		$1.33(2)$ 				&	$4/3$ 		& 	$8/5$	&	$1.77$	&	$5/3$\\
		$\alpha$	&		$8/5$					&		$1.67(2)$						&	$5/3$			&	$14/9$		&	$1.59(3)$		&	$23/15$
	\end{tabular}
	\caption{Critical exponents for Toner an Tu \cite{Toner:1998} and Vicsek \cite{Mahault:2019}, compared to the exact exponent computed in this work.}
	\label{table:exponents}
\end{table}

\vspace{0.8cm}
\section{Outlook} 
\noindent We have demonstrated that polar active fluids correspond to passive fluids lacking boost symmetry when the latter are supplemented with a non-thermal noise term and a potential inducing spontaneous symmetry breaking. In particular, thermodynamic relations must continue to hold in the ordered phase. Consequently, we precisely derived expressions for the critical exponent. These expressions align excellently with numerical simulations of the Vicsek model and also correlate well with experimental observations on epithelial cells \cite{Giavazzi:2017}, self-propelled rollers \cite{Geyer:2018}, bacterial colonies \cite{Zhang:2010}, polar rods \cite{Soni:2020} and earlier simulations of the Vicsek model \cite{Chate:2008,Ngo:2014}. This paves the way for more precise computations, where the constraint derived in our approach can give better agreement with experimental results.

We expect our argument to be valid for generic systems, like Malthusian flocks \cite{Toner:2012birth,chate2024} or incompressible phases \cite{Toner:1995}. It would also be interesting to show precisely how entropy conservation constrains the ideal-fluid transport coefficients at the level of the Feynman diagrams, even just for simple fluids without external potential $U$. Moreover, while we omitted the consideration of temperature, as is customary in the Toner-Tu model, it is straightforward to introduce an additional scalar quantity, \( T \), into our formalism. This leads to a conservation equation associated with time translation. Naturally, we anticipate this equation to undergo relaxation similar to that of the equation for \( \vec{g} \) \cite{Armas:2024iuy}. Understanding the implications of this relaxed equation could provide insight into the nature of entropy conservation/production within the effective active matter systems.

\begin{acknowledgements}
{\noindent The authors would like to thank John Toner for correspondence. A.A. \& D.B. have received support from the project PRIN 2022A8CJP3 by the Italian Ministry of University and Research (MUR). L.M. acknowledges support from the project PRIN 2022ZTPK4E by the Italian Ministry of University and Research (MUR). This project has also received funding from the European Union’s Horizon 2020 research and innovation programme under the Marie Sklodowska-Curie grant agreement No. 101030915.}
\end{acknowledgements}

\raggedright
\bibliography{refs}

\clearpage
\onecolumngrid
\appendix
\pagenumbering{alph}

\section{Constitutive relations for order-one boost-agnostic hydrodynamics}
\label{appendix:A}
\noindent In this appendix we give the constitutive relations for a boost agnostic fluid in $d$ spatial dimensions up to and including first order in derivatives. The constitutive relations for the current $j^i$ and $\tau^{ij}$ are \cite{Armas:2020mpr}:
\begin{subequations}
	\begin{align}
		j^i&=n v^i-\left(\gamma_{00}P^{ij}\partial_j\mu+\zeta_{00}\frac{v^iv^j}{\abs{v}^2}\partial_j\mu+\gamma_{01}\frac{P^{i(j}v^{k)}}{\abs{v}}\sigma_{jk}+\zeta_{01}\frac{v^iv^jv^k}{2\abs{v}^3}\sigma_{jk}+\frac{\zeta_{02}}{2}\frac{v^i}{\abs{v}}P^{jk}\sigma_{jk}\right)\ ,\\
		\tau^{ij}&=\rho v^i v^j+P\delta^{ij}-\left[\eta\left(P^{k(i}P^{j)l}-\frac{1}{d-1}P^{ij}P^{kl}\right)\sigma_{kl}+2\gamma_{01}\frac{P^{k(i}v^{j)}}{\abs{v}}\partial_k\mu+\zeta_{01}\frac{v^iv^jv^k}{\abs{v}^3}\partial_k\mu\right.\nonumber\\
		&\quad\left.+\zeta_{02}\frac{v^k}{\abs{v}}P^{ij}\partial_k\mu+2\gamma_{11}\frac{v^{(i}P^{j)(k}v^{l)}}{\abs{v}^2}\sigma_{kl}+\frac{1}{2}\zeta_{11}\frac{v^iv^jv^kv^l}{\abs{v}^4}\sigma_{kl}+\frac{\zeta_{12}}{2}\left(P^{ij}\frac{v^kv^l}{\abs{v}^2}+\frac{v^iv^j}{\abs{v}^2}P^{kl}\right)\sigma_{kl}\right.\nonumber\\
		&\quad\left.+\frac{\zeta_{22}}{2}P^{ij}P^{kl}\sigma_{kl}\right] \ ,
	\end{align}
\end{subequations}
where $\sigma_{ij}=\partial_iv_j+\partial_jv_i$, $P^{ij}=\delta^{ij}-\frac{v^iv^j}{\abs{v}^2}$ is the projector orthogonal to the velocity, and $\mu$ is the chemical potential. The momentum constitutive relation remains fixed at its ideal-fluid form, $\vec{g}=\rho\vec{v}$. Positivity of entropy production constrains the matrices of transport coefficients
\begin{equation}
	\begin{pmatrix}
		\zeta_{00}	&	\zeta_{01}	&	\zeta_{02}\\
		\zeta_{01}	&	\zeta_{11}	&	\zeta_{12}\\
		\zeta_{02}	&	\zeta_{12}	&	\zeta_{22}
	\end{pmatrix}\geq0 \ , \; \;
	\begin{pmatrix}
		\gamma_{00}	&	\gamma_{01}\\
		\gamma_{01}	&	\gamma_{11}
	\end{pmatrix}\geq0 \ , \;\; \eta\geq0 \ ,
\end{equation}
i.e. the $\zeta$ and $\gamma$ matrices are positive semi-definite. We can rewrite the constitutive relations using $n$ instead of $\mu$ as thermodynamic variable, which is more natural if we want to compare our results with \cite{Toner:1998}
\begin{subequations}\label{eqnSM:constitutive-relations}
	\begin{align}
		j^i&=n v^i-\left(\bar\gamma_{00}P^{ij}\partial_jn+\bar\zeta_{00}\frac{v^iv^j}{\abs{v}^2}\partial_jn+\bar\gamma_{01}\frac{P^{i(j}v^{k)}}{\abs{v}}\sigma_{jk}+\bar\zeta_{01}\frac{v^iv^jv^k}{2\abs{v}^3}\sigma_{jk}+\frac{\bar\zeta_{02}}{2}\frac{v^i}{\abs{v}}P^{jk}\sigma_{jk} + \mathfrak{m}P^{ij}\partial_jv^2\right)\ ,\\
		\tau^{ij}&=\rho v^i v^j+P\delta^{ij}-\left[\eta\left(P^{k(i}P^{j)l}-\frac{1}{d-1}P^{ij}P^{kl}\right)\sigma_{kl}+2\tilde\gamma_{01}\frac{P^{k(i}v^{j)}}{\abs{v}}\partial_kn+\tilde\zeta_{01}\frac{v^iv^jv^k}{\abs{v}^3}\partial_kn\right.\nonumber\\
		&\quad\left.+\tilde\zeta_{02}\frac{v^k}{\abs{v}}P^{ij}\partial_kn+2\tilde\gamma_{11}\frac{v^{(i}P^{j)(k}v^{l)}}{\abs{v}^2}\sigma_{kl}+\frac{1}{2}\tilde\zeta_{11}\frac{v^iv^jv^kv^l}{\abs{v}^4}\sigma_{kl}+\frac{\tilde\zeta_{12}}{2}\left(P^{ij}\frac{v^kv^l}{\abs{v}^2}+\frac{v^iv^j}{\abs{v}^2}P^{kl}\right)\sigma_{kl}\right.\nonumber\\
		&\quad\left.+\frac{\tilde\zeta_{22}}{2}P^{ij}P^{kl}\sigma_{kl}+2\frac{\mathfrak{t}}{\abs{v}}v^{(i}P^{j)k}\partial_kv^2+\frac{\mathfrak{n}}{\abs{v}}P^{ij}v^k\partial_kv^2\right] \ ,
	\end{align}
\end{subequations}
where we redefined the transport coefficients by absorbing the susceptibilities
\begin{subequations}
	\begin{align}
		\bar\gamma_{00}&=\gamma_{00}\left(\frac{\partial\mu}{\partial n}\right)_{v^2} \ ,	&	\bar\gamma_{01}&=\gamma_{01} \ ,	&	\bar\zeta_{00}&=\zeta_{00}\left(\frac{\partial\mu}{\partial n}\right)_{v^2} \ ,\\
		\bar\zeta_{01}&=\zeta_{01}+2\zeta_{00}\abs{\vec{v}}\left(\frac{\partial\mu}{\partial v^2}\right)_n \ ,	&	\bar\zeta_{02}&=\zeta_{02} \ ,	&	\mathfrak{m}&=\gamma_{00}\left(\frac{\partial\mu}{\partial v^2}\right)_n \ ,\\
		\tilde\gamma_{01}&=\gamma_{01}\left(\frac{\partial\mu}{\partial n}\right)_{v^2} \ ,	&	\tilde\gamma_{11}&=\gamma_{11} \ ,	&	\tilde\zeta_{01}&=\zeta_{01}\left(\frac{\partial\mu}{\partial n}\right)_{v^2}\ ,\\
		\tilde\zeta_{11}&=\zeta_{11}+2\zeta_{01}\abs{\vec{v}}\left(\frac{\partial\mu}{\partial v^2}\right)_n	\ ,&	\tilde\zeta_{02}&=\zeta_{02}\left(\frac{\partial\mu}{\partial n}\right)_{v^2}	&	\tilde\zeta_{12}&=\zeta_{12}\\
		\tilde\zeta_{22}&=\zeta_{22} \ ,	&		\mathfrak{t}&=\gamma_{01}\left(\frac{\partial\mu}{\partial v^2}\right)_n \ ,	&	\mathfrak{n}&=\zeta_{02}\left(\frac{\partial\mu}{\partial v^2}\right)_n \ .
	\end{align}
\end{subequations}
In spite of the fact that there are 10 independent order-one transport coefficients, the equations of motion linearized around a constant background velocity $\vec{v}=v_0\hat x_\parallel$ will be such that we can absorb all of them into three order-one diffusive coefficients in the charge conservation equation and four transport coefficients in the momentum equations.

\section{Transport coefficients from integrating out the longitudinal mode}
\label{appendix:B}
\noindent We introduce a notation, also used in the main text, for the thermodynamic derivatives of the external potential $U$, the pressure $P$, the kinetic mass density $\rho$ and the entropy density $s$
\begin{align}\label{eqnSM:notation}
	\left(\frac{\partial F}{\partial n}\right)_{v^2}&=F_n \ ,	&	\left(\frac{\partial F}{\partial\vec{v}^2}\right)_n&=F_v	\ ,&	\left(\frac{\partial^2 F}{\partial n^2}\right)_{v^2}&=F_{n^2} \ ,	&	\left(\frac{\partial^2 F}{\partial(\vec{v}^2)^2}\right)_{n}&=F_{v^2} \ ,	&	\frac{\partial^2 F}{\partial n\partial\vec{v}^2}&=F_{n,v}
\end{align}
with $F=\{U,P,\rho,s\}$. These thermodynamic derivatives are evaluated on the background, at $n=n_0$ and $\vec{v}^2=v_0^2$. We remark that, without loss of generality, we set $U_{v^2}=U_{n,v}=0$. Reinstating these derivatives simply changes the values of $\alpha_i$ in \eqref{eqn:frozen_profile}, but does not affect any conclusion.

In the main text, when we define the effective equation for fluctuations of the charge density once the longitudinal mode is integrated out, we make use of the following definitions
\begin{subequations}\label{eqnSM:charge-transport-coefficients}
    \begin{align}
	v_2&=v_0-\alpha_1n_0 \ ,	&	w_1&=1 \ ,\\
	w_2&=\alpha_1+\alpha_2n_0\ ,	&	w_3&=\frac{n_0}{2v_0} \ ,\\
	\xi_n&=\alpha_3n_0+\bar\gamma_{01}+\bar\zeta_{02}\ ,	&	D_\parallel&=n_0\alpha_4+\bar\zeta_{00}-\alpha_1\bar\zeta_{01} \ ,\\
	D_\perp&=\bar\gamma_{00}-\alpha_1(2v_0\mathfrak{m}+\bar\gamma_{01}) \ .
\end{align}
\end{subequations}
Similarly, for the transverse velocity fluctuations we find
\begin{subequations}\label{eqnSM:momentum-transport-coefficients}
	\begin{align}
		\kappa&=\frac{1}{\rho_0}\left(P_n-2v_0\alpha_1P_v\right) \ ,	&		g_1&=-\alpha_1 \ ,\\
		g_2&=\frac{1}{\rho_0}\left(n_0\alpha_1\rho_n-2\alpha_1^2n_0v_0\rho_v+2v_0\alpha_4U_v\right)-\alpha_1 \ ,\\
		g_3&=\frac{P_{n^2}}{2\rho_0}+\frac{P_v}{\rho_0}\left(\alpha_1^2-2v_0\alpha_2\right)+\frac{\kappa\left(2v_0\alpha_1\rho_v-\rho_n\right)}{2\rho_0}-\frac{2v_0\alpha_1P_{n,v}}{\rho_0}+\frac{2v_0^2\alpha_1^2P_{v^2}}{\rho_0} \ ,\\
		g_4&=1-\frac{1}{\rho_0}\left(n_0\rho_n-2v_0n_0\alpha_1\rho_v-2v_0\alpha_3U_v\right) \ ,	&		g_5&=1 \ ,\\
		D_v&=\frac{1}{\rho_0}\left(2v_0\alpha_4P_v+\tilde\gamma_{01}+\tilde\zeta_{02}-\alpha_1(2v_0(\mathfrak{t}+\mathfrak{n})+\tilde\gamma_{11}+\tilde\zeta_{12})\right) \ ,	&	\xi_v&=\frac{1}{\rho_0}\left(2v_0\alpha_3P_v+\tilde\zeta_{22}\right) \ ,\\
		\xi_\parallel&=\frac{\tilde\gamma_{11}}{\rho_0} \ ,	&		\xi_\perp&=\frac{\eta}{\rho_0} \ .
	\end{align}
\end{subequations}
Finally, the coefficients that appear in the entropy equation \eqref{eqn:entropy_reduced} are
\begin{subequations}
	\begin{align}
		\varphi&=s_0T_0-P_0+2\alpha_3v_0^3U_v \ ,\\
		\pi&=v_0\left(T_0s_n-P_n\right)+\alpha_1\left(P_0-s_0T_0+2v_0^2(P_v-T_0s_v	)\right)+2\alpha_4v_0^3U_v \ ,\\
		f_1&=T_0s_n-P_n+v_0\left(2\alpha_1P_v-2\alpha_1T_0s_v-3\alpha_3U_n\right) \ ,\\
		f_2&=T_0s_n-P_n+2v_0\alpha_1\left(P_v-T_0s_v\right) \ ,\\
		f_3&=\frac{P_0-s_0T_0}{2v_0} \ ,\\
		f_4&=\frac{v_0\left(T_0 s_{n^2}-P_{n^2}\right)}{2}+\alpha_1\left(P_n-T_0s_n\right)+2v_0^2\alpha_1\left(P_{n,v}-T_0s_{n,v}\right)-3\alpha_1^2v_0\left(P_v-T_0s_v\right)\nonumber\\
		&\qquad\qquad\qquad-2v_0^3\alpha_1^2\left(P_{v^2}-T_0s_{v^2}\right)+\alpha_2\left(P_0-T_0s_0\right)+2v_0\alpha_2\left(P_v-T_0s_v\right)-\frac{3}{2}\alpha_4v_0U_n \ .
	\end{align}
\end{subequations}

\end{document}